\documentclass[12pt,twoside]{article}
\usepackage{fleqn,espcrc1}

\usepackage{epsf}

\newcommand{\AmS}{{\protect\the\textfont2
  A\kern-.1667em\lower.5ex\hbox{M}\kern-.125emS}}

\title{
\vspace*{-60pt}
{\normalsize \hfill {\sf UTHEP-425}} \\
\vspace*{-5pt}
{\normalsize \hfill {\sf UTCCP-P-86}} \\
\vspace*{-5pt}
{\normalsize \hfill {\sf May 2000}} \\
\vspace*{30pt}
	Hadronic Properties from Lattice QCD with Dynamical Quarks
\footnote{Talk presented at the 
International Conference on Quark Nuclear Physics (QNP2000),
21-25 February, 2000, Adelaide, Australia.}
}

\author{K. Kanaya%
\address{Institute of Physics, University of Tsukuba, 
	Tsukuba, Ibaraki 305-8571, Japan}
}

\begin{document}
\maketitle

\begin{abstract}
The lattice regularization of QCD provides us with the most systematic
way of computing non-perturbative properties of hadrons directly from
the first principles of QCD.
The recent rapid development of parallel computers has enabled us to start
realistic and systematic simulations with dynamical quarks. 
In this paper, I report on the first results from recent systematic 
studies on the lattice with dynamical quarks.
\end{abstract}

\section{INTRODUCTION}
\label{sec:intro}

We formulate QCD on a 4-dimensional hyper-cubic lattice with a
finite lattice spacing $a$. 
When the lattice volume is also finite, the theory is completely finite 
and well-defined, so that supercomputers can be used to calculate
non-perturbative quantities.
Numerous developments and ideas in the last two decades, both 
in algorithms for numerical computations and in the computer technology, 
have made lattice field theory one of the most powerful tools in computing
the non-perturbative properties of QCD.
In particular, the development of parallel computers in the late 90's
enabled us to start realistic and systematic simulations of QCD. 

In order to extract predictions for the real world from the numerical 
results obtained on finite lattices, the following extrapolations are required:
Because the continuum physics is defined in the limit of large lattice
volume and vanishing lattice spacing, it is necessary to
\begin{enumerate}
\item\hspace*{-2mm}) extrapolate to vanishing lattice spacing,
\item\hspace*{-2mm}) extrapolate to large lattice volume.
\end{enumerate}
Here,
computations for the quark part contain inversions of a huge quark matrix, 
whose size is, for example, about $10^7\times 10^7$ for the case of a 
$32^4$ lattice. 
Because the condition number of the quark matrix diverges as we decrease 
the quark mass, a calculation with light quarks is quite time-consuming. 
Actually, even with the latest supercomputers with TFLOPS level computational 
speeds, a direct simulation with the physical $u$ and $d$ quarks is not 
possible. 
Therefore, at present, we also have to 
\begin{enumerate}
\addtocounter{enumi}{+2}
\item\hspace*{-2mm}) extrapolate to light $u$ and $d$ quarks, 
\end{enumerate}
using data at around the $s$ quark mass region. 

To obtain a reliable result for continuum physics, it is essential 
to have a good control of the systematic errors due to these 
extrapolations. 
A systematic approach is required.
Because of the huge amount of computations required, 
major calculations have been made in the quenched approximation, 
in which the effects of dynamical quark loops are ignored. 
A recent extensive study of quenched QCD by the CP-PACS Collaboration 
\cite{CPPACSquench}, however, has shown a limitation of the quenched 
approximation. 
As a next logical step, systematic ``full QCD'' simulations removing
the quenched approximation have started.
In this paper, I summarize the latest status of lattice QCD simulations, 
choosing ``dynamical quark effects'' as the keyword.

In Secs.~\ref{sec:spectrum} and \ref{sec:fspectrum}, the latest results
for the light hadron spectrum from lattice QCD are summarized. 
This provides us with the most direct test of QCD in the low-energy 
non-perturbative domain.
In Sec.~\ref{sec:Mq}, lattice results for the mass of light quarks are 
discussed. 
Sec.~\ref{sec:U1} is devoted to the status of the U(1) problem
on the lattice.
In Sec.~\ref{sec:fB}, lattice results for $B$ meson decay constants are
summarized.
We find that, in all of these physics, dynamical quarks have 
significant effects.
A short conclusion is given in Sec.~\ref{sec:conclusion}.

\section{QUENCHED LIGHT HADRON SPECTRUM}
\label{sec:spectrum}

\begin{figure}[bt]
\centerline{\epsfxsize=9cm \epsfbox{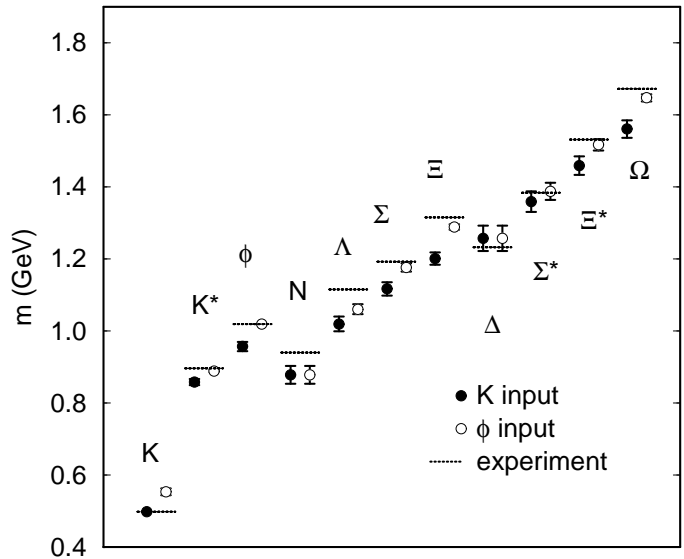}}
\vspace{-5mm}
\caption{Quenched light hadron spectrum with 
degenerate $u$ and $d$ quarks \cite{CPPACSquench}.}
\label{fig:qspectrum}
\end{figure}

The computation of the hadronic mass spectrum directly from the first 
principles of QCD is one of the most important tasks of lattice QCD
\cite{review,review2}. 
If the lattice spectrum reproduces the experimental data, this will
provide us with the most convincing demonstration of the validity of 
QCD as the fundamental theory for strong interactions in the low-energy 
non-perturbative domain.
Furthermore, through these calculations, we can see how the lattice QCD
describes the real physics and what is required to achieve high accuracy 
suppressing lattice artifacts. 
Another motivation is a determination of fundamental parameters in QCD;
see Sec.~\ref{sec:Mq}.
 
Because huge computer power is required to overcome the systematic 
errors discussed in Sec~\ref{sec:intro}, major 
simulations have been done in the quenched approximation.
With this approximation, the amount of computations is reduced by a 
factor of several hundreds.

Recently, the CP-PACS Collaboration performed a large-scale calculation of
this with a precision significantly better than previous studies
\cite{CPPACSquench}.
Using a dedicated massively parallel computer CP-PACS developed 
at the University of Tsukuba \cite{cp-pacs}, they studied four lattices 
$32^3\times56$ to $64^3\times112$ with lattice spacing in the range 
$a\approx 0.1$--0.05 fm.
The spacial lattice size was fixed to be about 3 fm with which the finite
size effects are estimated to be maximally 0.5\% in the spectrum.
On each lattice, five quark masses, corresponding to the mass ratio
of pseudo-scalar and vector mesons $m_{\rm PS}/m_{\rm V}\approx 0.75$--0.4,
were studied.
The light $u$ and $d$ quarks were treated as degenerate.
Their final results in the continuum limit are summarized in 
Fig.~\ref{fig:qspectrum}.
The light quark mass $m_{ud}$ and the lattice scale $a$ were fixed using
the experimental values for $m_\pi$ and $m_\rho$ as inputs, while the
$s$ quark mass was fixed either by $m_K$ ($K$-input) or $m_\phi$ 
($\phi$-input).
See \cite{CPPACSquench} for details of the simulation and extrapolations.

From Fig.~\ref{fig:qspectrum}, we see that, although the global pattern 
of the experimental spectrum is reproduced, there remain systematic 
discrepancies of up to about 10\% (7 standard deviations) 
between quenched QCD and experiment, for both choices of the input
for the strange quark mass.
Because other systematic errors are well under control, these 
discrepancies might be due to the quenched approximation.

\section{FULL QCD SPECTRUM}
\label{sec:fspectrum}

In the previous section, a limitation of the quenched approximation 
was made clear. 
The next logical step is to perform a ``full QCD'' calculation removing
the quenched approximation, to see if these discrepancies disappear
with dynamical quarks. 
However, a naive extension of quenched simulations is difficult
as the computational power required increases several hundred fold. 
Recently, there was much progress in improving lattice actions, with 
which lattice artifacts due to finite lattice spacing are reduced.
Therefore, a coarser lattice may be used for continuum extrapolations
with correspondingly less computer time.

\begin{figure}[t]
\vspace{-5mm}
\begin{minipage}[t]{75mm}
\centerline{\epsfxsize=7.5cm \epsfbox{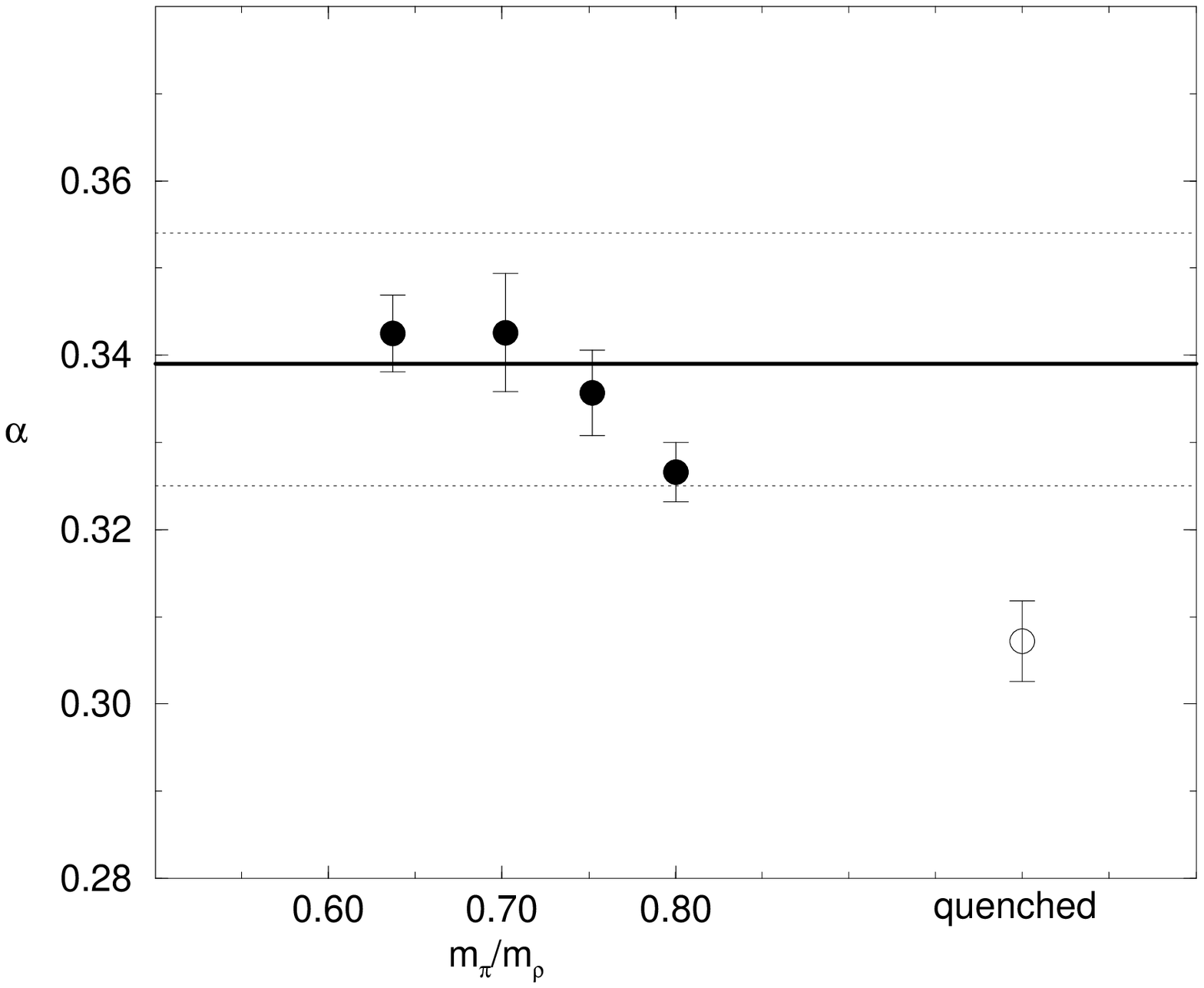}}
\vspace{-10mm}
\caption{Coulomb coefficient in full (filled circle) and 
quenched QCD (open circle).
Solid line represents an estimate expected for the full QCD value
in the chiral limit, and dashed lines its error.}
\label{fig:alpha}
\end{minipage}
\hspace{\fill}
\begin{minipage}[t]{80mm}
\centerline{\epsfxsize=8cm \epsfbox{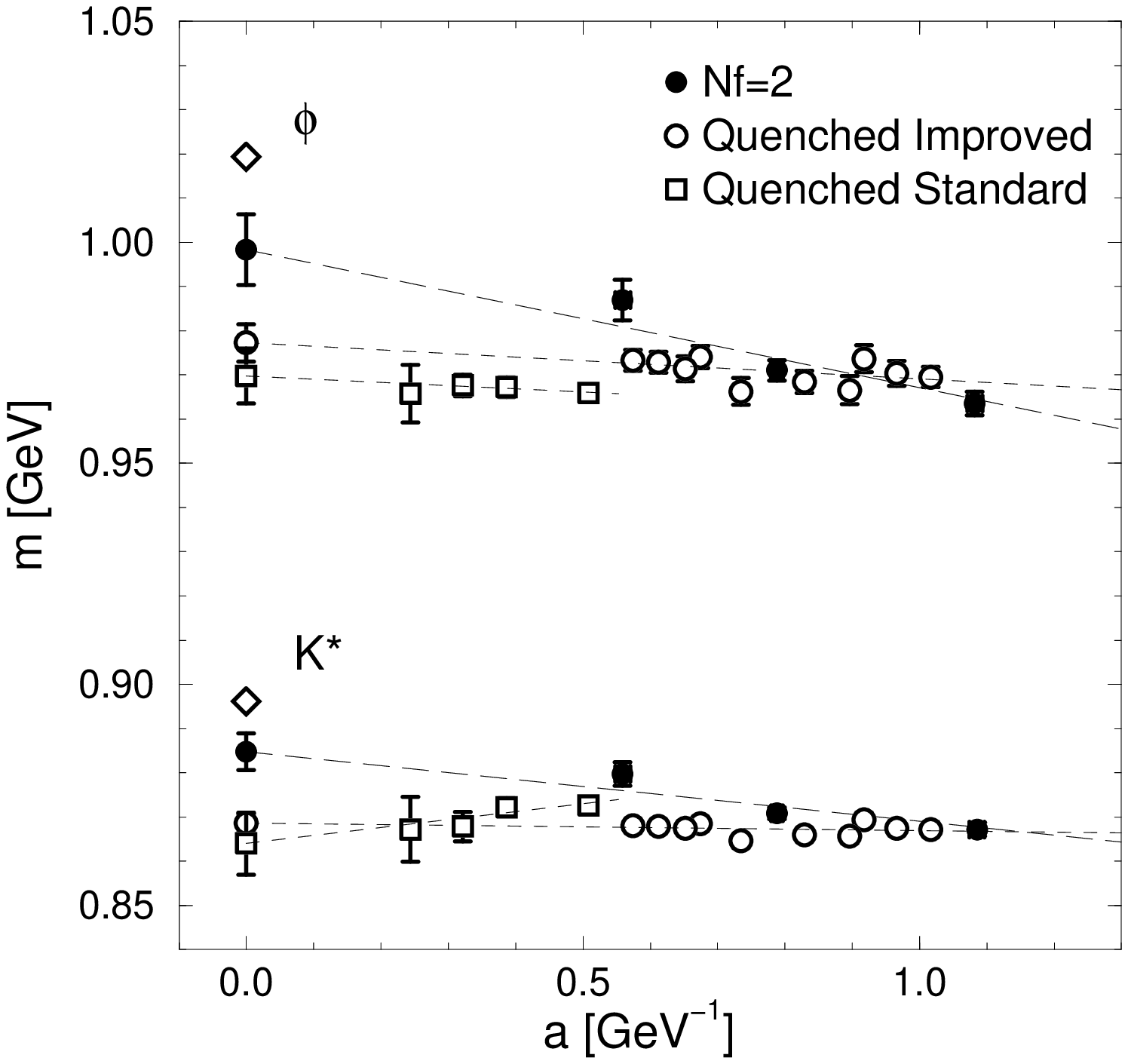}}
\vspace{-10mm}
\caption{Continuum extrapolation of vector meson masses $m_{\phi}$ and
$m_{K^*}$ in full and quenched QCD, using the $K$ meson mass as input
\cite{CPPACSfull}.}
\label{fig:fspectrum}
\end{minipage}
\end{figure}

The CP-PACS Collaboration adopted the combination of an RG-improved 
gauge action and a ``clover''-type improved Wilson quark action, 
and carried out the first systematic investigation of full QCD.
Two flavors of dynamical quarks, to be identified with degenerate $u$ 
and $d$ quarks, are simulated \cite{CPPACSfull}. 
The heavier $s$ quark was treated in the quenched approximation.

We may ask ourselves if we can actually see an effect of dynamical 
quarks on the present lattice. The most direct signals of dynamical quarks 
will be decays, such as $\rho \rightarrow \pi \pi$, and the breaking
of the confining string for the static quark potential. It turned out that
these phenomena are inaccesible from the current simulation parameters 
(quark mass, lattice size, etc.) \cite{review2}. 
We can, however, see an effect of dynamical quarks in the short range physics: 
The phenomenological string tension $\sigma \simeq (440 {\rm MeV})^2$ 
from the charmonium spectrum, is considered to represent the physics at 
a scale $\mu \approx 400$--600 MeV.
Therefore, when we compare the quenched and full QCD at the same 
lattice string tension,  
we are connecting the two theories at $\mu \approx 400$--600 MeV.
Because the running of effective coupling is different between quenched 
and full QCD,
we should see a difference in the Coulomb coefficient $\alpha$ 
which represents the physics at $\mu \sim 1$--2 GeV \cite{shift.of.alpha}.
In Fig.~\ref{fig:alpha}, results for $\alpha$ obtained on a 
$24^3\times 48$ lattice are shown for full and quenched QCD 
at the same lattice string tension \cite{CPPACSfull}.
As expected, the full QCD results are larger than the quenched value
at $m_{\rm PS}/m_{\rm V}=\infty$, 
and increase as the sea quark becomes lighter.
The horizontal line in Fig.~\ref{fig:alpha} is an expected value 
for $\alpha$ at $m_{\rm PS}/m_{\rm V}=0$ in full QCD, 
estimated from the quenched value using the
two-loop beta function with two flavors of massless quarks.
The full QCD values for light sea quarks are consistent with 
this estimate. 
Therefore, we do have dynamical quark effects with the expected 
magnitude.

Now, we study implications of dynamical quarks 
in non-perturbative physics.
Figure~\ref{fig:fspectrum} shows the lattice spacing dependence
of hadron masses from $N_F=2$ full QCD, compared with quenched 
masses. Plotted together are the results of a quenched calculation
using the same improved action as in the full QCD calculation, with 
which one-to-one comparisons between quenched and full QCD results 
from calculations with similar parameters are made possible.
We see that both quenched calculations lead to universal values
for hadron spectrum in the continuum limit, and the discrepancies
from experiment in quenched spectrum are much reduced by two flavors 
of dynamical quarks.
The remaining difference might be caused by the quenching of the $s$
quark. 

\section{QUARK MASSES}
\label{sec:Mq}

Massses of quarks belong to the most fundamental parameters of QCD. 
Because quarks are confined in hadrons, their masses have
to be indirectly inferred from hadronic observables using the 
functional relation between these hadronic quantities and QCD 
parameters. 
A lattice QCD determination of the hadron spectrum provides us with 
such functional relations without resorting to a phenomenological
model. 

\begin{figure}[t]
\centerline{
\epsfxsize=80mm \epsfbox{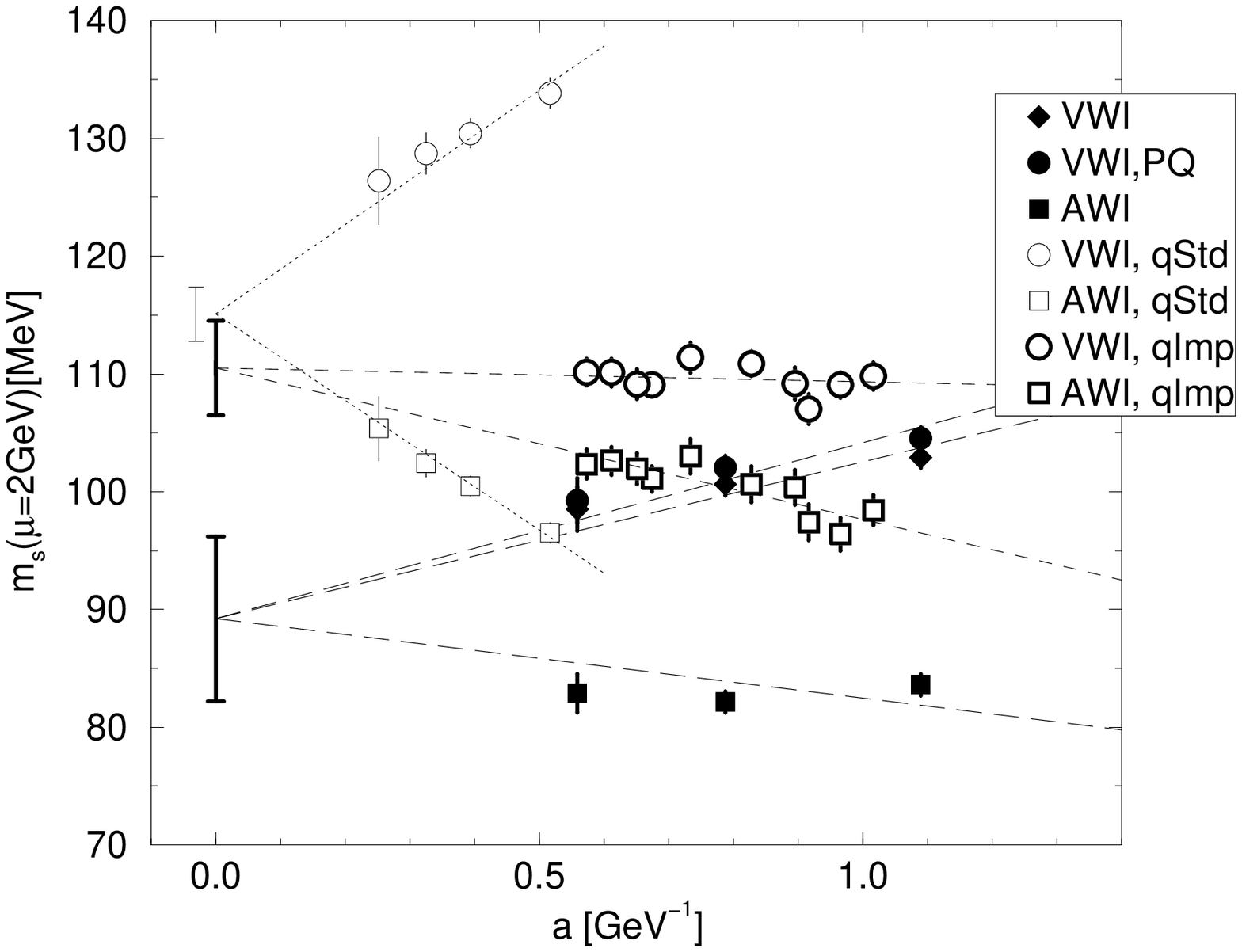}
\epsfxsize=80mm \epsfbox{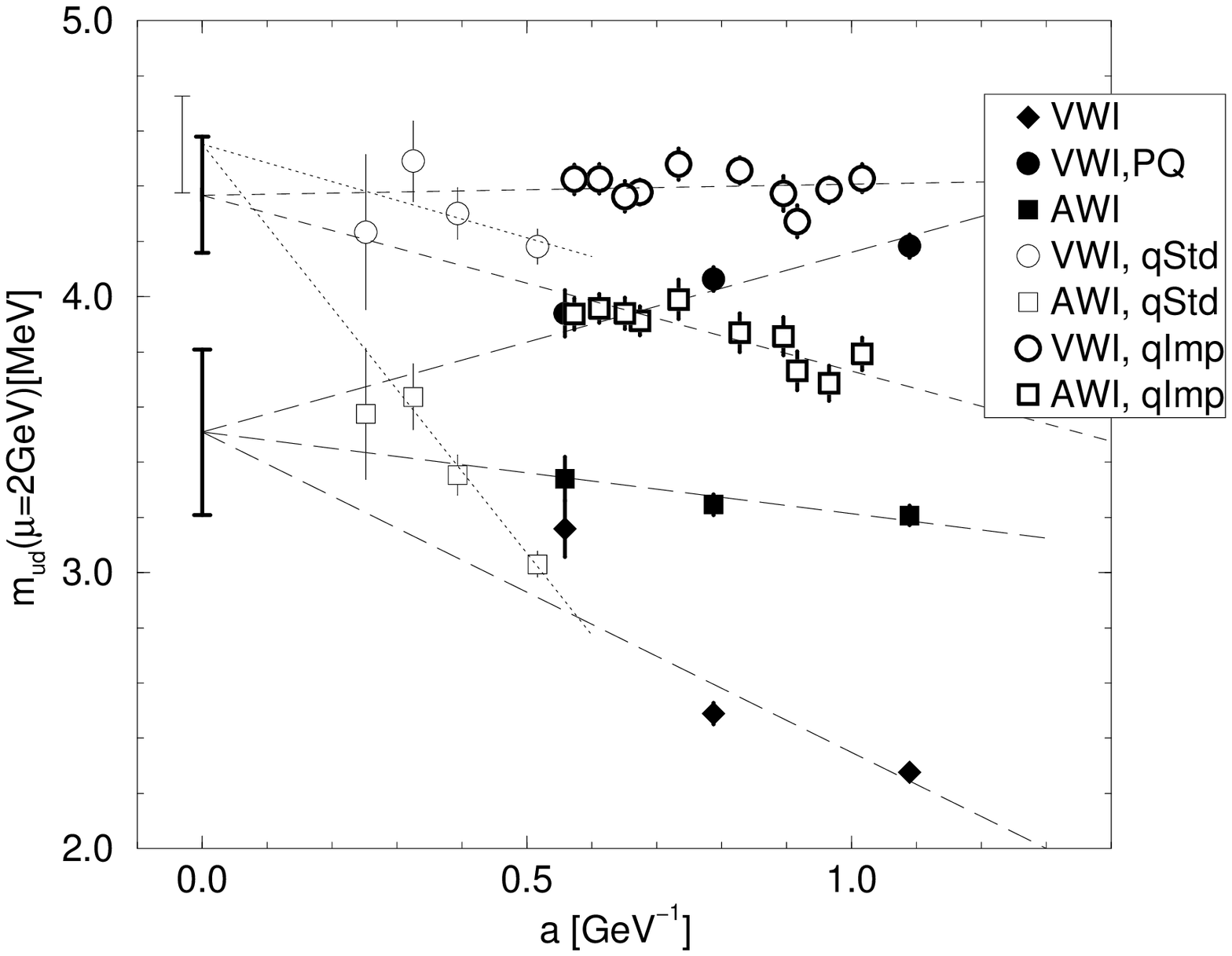}
}
\vspace{-10mm}
\caption{Continuum extrapolation of the average $u$ and $d$ quark mass
$m_{ud}$ and the $s$ quark mass $m_s$ in the $\overline{\rm MS}$
scheme at 2 GeV. $m_s$ is from the $K$-input.
Filled symbols are for full QCD. Quenched results with the standard action
and the improved action are shown with thin and thick open symbols,
respectively.}
\label{fig:ms}
\end{figure}

\begin{table}[b]
\caption{Results for quark masses in the $\overline{\rm MS}$
scheme at 2 GeV}
\begin{tabular}{lccc}
    &  $m_{ud}$ (MeV)  & $m_s$ (MeV) ($K$-input) & 
       $m_s$ (MeV) ($\phi$-input)  \\
\hline
$N_f=0$ standard & 4.55(18) & 115(2) & 143(6) \\
$N_f=0$ improved & 4.4(2)   & 110(4) & 132(3) \\
$N_f=2$          & 3.5(3)   &  89(7) & 99(5)  \\
\hline
\end{tabular}
\label{tab:quarkmass}
\end{table}

On the lattice, there exist several alternative definitions for quark 
mass. 
They should converge to a universal value in the continuum limit.
The CP-PACS Collaboration observed that 
light quark masses determined from either the vector Ward identity
(VWI) or the axial-vector Ward identity (AWI), while differing at finite
lattice spacings, actually converge to a common value in the continuum 
limit \cite{CPPACSquench}. See Fig.~\ref{fig:ms}. 
However, in the quenched QCD, the resulting value for $m_s$ 
differs by about 20\% between $K$-input and $\phi$-input, 
due to the discrepancy of quenched spectrum from experiment discussed 
in Sec.~\ref{sec:spectrum}. 

In Fig.~\ref{fig:ms}, $N_F=2$ full QCD values, as well as quenched
values with the improved action, are also plotted \cite{CPPACSmq}.
We again find that the two quenched simulations lead to a universal value
in the continuum limit.
Results for the light quark masses in the continuum limit are summarized 
in Table \ref{tab:quarkmass}. 
The discrepancy between the $K$ and $\phi$-inputs becomes much smaller
in full QCD. 
The most impressive point is that the values predicted through full
QCD are 20--30\% smaller than those in the quenched QCD. In particular,
the $s$ quark mass in two-flavor full QCD is about 90--100 MeV, 
which is significantly smaller than the phenomenological value $\approx 
150$ MeV often used in hadron phenomenology, 
and almost saturating an estimate of the lower bound 90--100 MeV 
from QCD sum rules \cite{narison99}.

\section{U(1) PROBLEM}
\label{sec:U1}

Another important issue in QCD is the $U(1)$ problem, that is to 
clarify the mechanism for a large $\eta^\prime$ meson mass.
Propagators of flavor non-singlet meson consist of a loop of 
valence quark propagator, while propagators of flavor singlet 
$\eta^\prime$  meson have an additional contribution 
with two disconnected valence quark loops.
The fact that $\eta'$ is much heavier than the corresponding 
non-singlet meson $\pi$ means that the two-loop contribution 
should exactly cancel the $\pi$ pole of the one-loop contribution, 
leaving the heavy $\eta'$ pole.
This phenomenon is considered to be related with the anomalous 
violation of the flavor singlet axial U(1) symmetry and 
topological structure of gauge field configurations in QCD.
This is a critical test of QCD and can be answered using lattice QCD.

Calculation of the two-loop contribution requires a large amount of 
computations.  For this reason only limited results have been available.
The first systematic investigation in full QCD including continuum 
extrapolations was recently done by the CP-PACS Collaboration 
\cite{CPPACSfull}.
In an approximation ignoring with the $s\bar{s}$ state, 
the flavor-singlet $u\bar{u}+d\bar{d}$ meson mass was estimated as 
$m_{u\bar{u}+d\bar{d}}=863\pm86$~MeV. 
In the real world, the $u\bar{u}+d\bar{d}$ state mixes with the 
$s\bar{s}$ state to lead to $\eta$(547) and $\eta'$(958) mesons.
Therefore, the result $863\pm86$~MeV which is slightly smaller than 
the experimental value 958~MeV is quite encouraging.  
Further studies, including an inspection of the mixing with the 
$s\bar{s}$ state as well as the study of topological structures, 
are under way.

\section{B MESON DECAY CONSTANTS}
\label{sec:fB}

The study of the non-perturbative properties of QCD is important also 
to investigate the Weinberg-Salam theory for the electro-weak 
interactions \cite{buras99}.
In order to determine CKM parameters for quarks, we need 
to know QCD corrections (hadronic matrix elements, form factors, etc.) 
in weak decays and mixings. 
In particular, properties of heavy mesons consisting of a $b$ quark 
are being studied intensively by the B factories. 
The precise determination of the decay constants and $B$-parameters 
for $B (= B_d)$ and $B_s$ mesons will lead to a strong constraint on 
the value of CKM parameters.
The lattice calculation of these quantities are beginning to contribute 
in reducing the ambiguities in these studies \cite{hashimoto99}.

Major calculations have been done in the quenched approximation.
Recently, full QCD calculations of the decay constants $f_B$ and 
$f_{B_s}$ have been made by several groups.
In this section, I concentrate on the dynamical quark effects in 
these decay constants.

On the lattice, simulation of the heavy $b$ quark is not a trivial
extension of light quark simulations, because $m_b \sim 4$ GeV is 
larger than the lattice cutoff $\sim 1$--4 GeV to date. 
Two methods have been developed to simulate heavy quarks directly.
One is based on a non-relativistic effective theory of QCD 
(NRQCD) defined through an expansion in the inverse heavy quark mass
\cite{NRQCD}.
Another employs a relativistic action and reinterpret it in terms
of a non-relativistic Hamiltonian \cite{Fermilab}.

\begin{figure}[bt]
\begin{minipage}[t]{80mm}
\centerline{\epsfysize=5cm \epsfbox{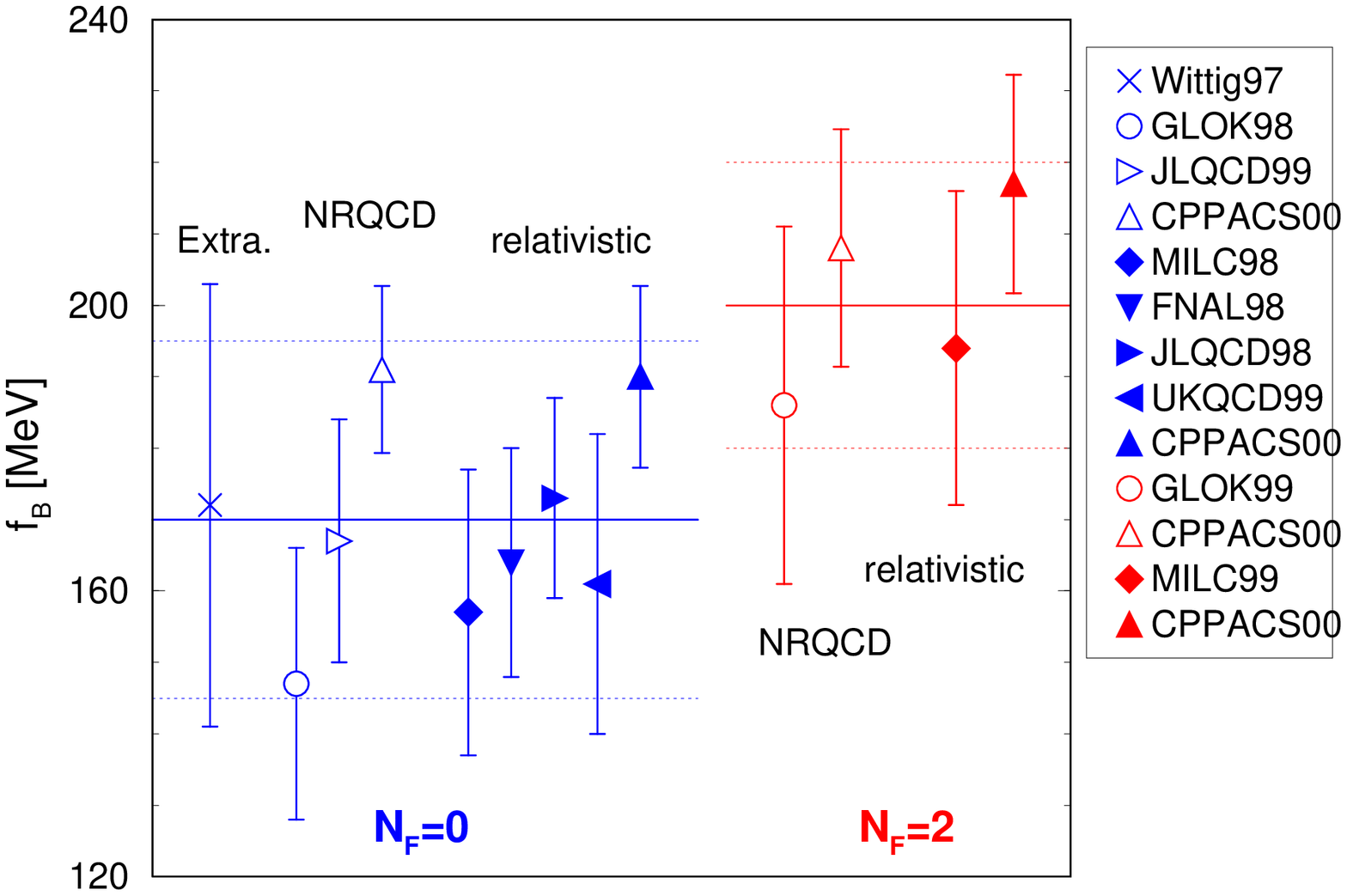}}
\vspace{-5mm}
\caption{Recent lattice results for the $B$ meson decay constant
from quenched simulations ($N_F=0$) and two-flavor full QCD 
simulations ($N_F=2$), with various methods.
}
\label{fig:fball}
\end{minipage}
\hspace{\fill}
\begin{minipage}[t]{75mm}
\centerline{\epsfysize=6.5cm \epsfbox{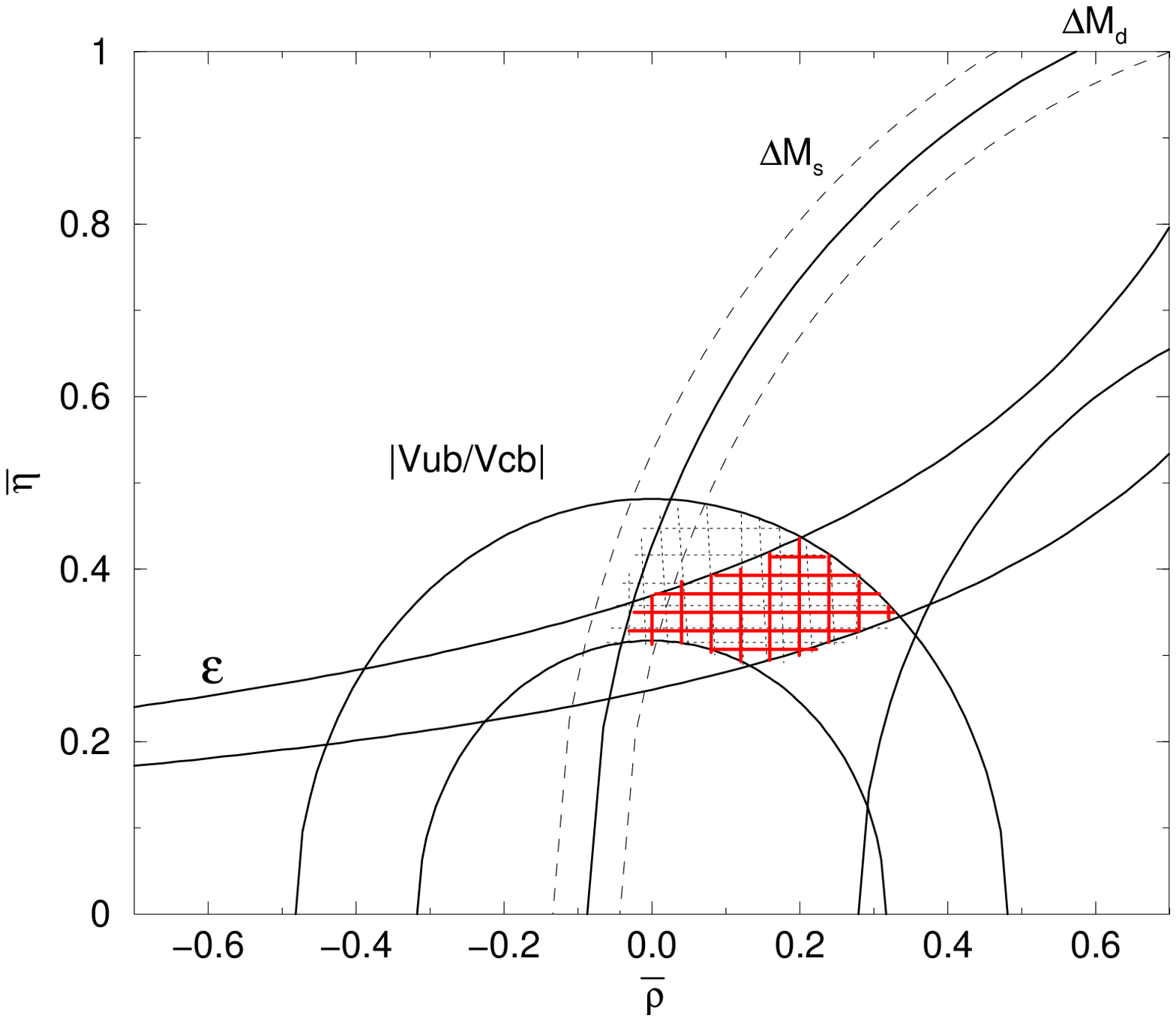}}
\vspace{-5mm}
\caption{Constraints on the CKM parameters $\bar\rho$ and $\bar\eta$
using lattice results. 
Allowed region from all the constraints is shaded by thick lines.
See text for details.
}
\label{fig:rhoeta}
\end{minipage}
\end{figure}

In Fig.~\ref{fig:fball}, I collect recent lattice results for $f_B$
from quenched 
\cite{Wittig97,GLOK98,JLQCD99,MILC98,FNAL98,JLQCD98,UKQCD99,CPPACSfB}
and full QCD studies \cite{CPPACSfB,GLOK99,MILC99}.
In the figure, ``Extrap.'' is a result of an extrapolation from light
quark mass region, and ``NRQCD'' and ``relativistic'' are the results
of the lattice NRQCD method and the relativistic method. 
Estimated statistical and systematic errors are included.
Although the errors are still large, 
we can see that both methods are approximately consistent with each other.
Comparing the quenched results ($N_f=0$) and the full QCD $N_f=2$ results, 
the latters seem to be about 10\% larger than the former values.
From this figure, I estimate $f_B = 200 \pm 20$ MeV.
Also for the case of $f_{B_s}$, increase of a similar magnitude is observed.

In Fig.~\ref{fig:rhoeta},
the implication of the lattice results for $f_B$ and $f_{B_s}$ 
on the unitarity triangle of CKM parameters is summarized
for 
$\bar\rho = \rho (1-\lambda^2/2)$ and 
$\bar\eta = \eta (1-\lambda^2/2)$
using the Wolfenstein parameterization (see \cite{buras99} for details).
Regions surrounded by two arcs are the allowed regions from each 
constraints, except for the two thin dashed lines denoted as $\Delta M_s$
which show the range of lower bounds for $\bar\rho$.
In this figure, the quenched lattice results for $B_K$, 
$B_B$, etc.\ are used in the constraints from $\epsilon$ and 
$\left| V_{ub}/V_{cb}\right|$ \cite{aoki:lp99}, 
while two-flavor full QCD results for $f_B$ and $f_{B_s}$ are used 
for the constraints from $\Delta M_d$ and $\Delta M_s$. 
Allowed region from all the constraints is shaded by thick lines.
The increase of $f_B$ due to the dynamical light quarks causes a shift 
of the allowed region from $\Delta M_d$ towards positive $\bar{\rho}$,
leading to a more severe constraint.
The region shaded by thin lines is the allowed region from 
the conventional phenomenological estimates for hadronic matrix 
elements \cite{buras99,aoki:lp99}.
Hence, lattice results are contributing to reduce errors in the 
determination of CKM parameters.

\section{CONCLUSIONS}
\label{sec:conclusion}

Due to the large computer power required, major lattice simulations
have been made in the quenched approximation, in which the dynamical
sea quark effects are ignored. 
Recently, it became possible to start realistic and systematic 
simulations of QCD with dynamical quarks, through the development of
parallel computers.
From the first simulations, it turned out that dynamical quark effects 
are quite important.
The effect of two flavors of dynamical light $u$ and $d$ quarks 
is as large as about 20\% in the values of light quark mass
and about 10\% in $B$ meson decay constants. 
Both of the shifts has significant implications to phenomenological 
studies of the standard model.

I thank the members of 
the CP-PACS Collaboration \cite{CPPACScollab} for discussions. 
In particular, I am grateful to Hugh Shanahan for useful comments 
and preparation of several figures.
I also thank Tony Thomas and other participants of the Conference 
for stimulative discussions at Adelaide.
This paper is in part supported by 
the Grants-in-Aid of Ministry of Education, Science and Culture 
(No.~10640248) and JSPS Research for Future Program.


\end{document}